\documentclass[12pt,journal]{IEEEtran}
\usepackage{url}
\usepackage{amsmath,graphicx,cite}
\usepackage{amsfonts}
\usepackage{algpseudocode}
\usepackage[ruled,linesnumbered]{algorithm2e}
\usepackage{verbatim}
\usepackage{longtable}
\usepackage{rotating}
\usepackage{float}
\usepackage{multirow}
\usepackage{tabularx}
\usepackage{bigints}
\usepackage{tikz}
\usepackage{multirow} 
\usepackage{amsmath,amstext,amsfonts,amssymb}
\usepackage{amsbsy}
\usepackage{amsthm}
\usepackage{diagbox}
\usepackage{mathabx}
\usepackage{mathrsfs}
\usepackage{bbm}
\usepackage{color}
\usepackage{caption}
\usepackage{subcaption}
\usepackage{morefloats}
\usepackage{float}
\usetikzlibrary{arrows,shapes,positioning,shadows,trees}
\usetikzlibrary{decorations.pathreplacing}

\begin{document}

\title{\textcolor{black}{Goal-Oriented Communications for the IoT and Application to Data Compression}}

\author{Chao~Zhang, Hang~Zou,~Samson~Lasaulce,~Walid~Saad, Marios~Kountouris, and Mehdi~Bennis
\thanks{This work is recently accepted by IEEE IoT magazine.}
\thanks{This work was supported in part by the RTE-CentraleSupelec Chair, and also by the Office of Naval Research (ONR) MURI Grant  N00014-19-1-2621. In addition, M. Kountouris has received funding from the European Research Council (ERC) under the European Union’s Horizon 2020 research and innovation programme (Grant agreement No. 101003431).}
\thanks{Chao Zhang is  with School of Computer Science and Engineering, Central South University, Changsha 410083, China.  Hang Zou is the corresponding author (Hang.Zou@tii.ae) with Technology Innovation Institute, Masdar City, Abu Dhabi, United Arab Emirates. Samson Lasaulce is with the Research Center for Automatic Control (CRAN) of CNRS and University of Lorraine at Nancy, France. Walid Saad is with Wireless @ VT, Department of Electrical and Computer Engineering, Virginia Tech, Blacksburg, VA 24061
USA. Marios Kountouris is with the Communication Systems Department, EURECOM, 06410 Biot, France. Mehdi Bennis is with Centre for Wireless Communications, University of Oulu, Finland.}}



\maketitle

\begin{abstract}
Internet of Things (IoT) devices will play an important role in emerging applications, since their sensing, actuation, processing, and wireless communication capabilities stimulate data collection, transmission and decision processes of smart applications. However, new challenges arise from the widespread popularity of IoT devices, including \textcolor{black}{the need for processing} more complicated data structures and high dimensional data/signals. \textcolor{black}{The unprecedented volume, heterogeneity, and velocity of IoT data calls for a communication paradigm shift from \textcolor{black}{a search for accuracy or fidelity to semantics extraction and} goal accomplishment.} In this paper, we provide \textcolor{black}{a partial but insightful overview} of recent research efforts in \textcolor{black}{this newly formed} area of goal-oriented (GO) and semantic communications, focusing on the problem of GO data compression for IoT applications. 
\end{abstract}


\section{Introduction}

The term Internet of Things (IoT) generally refers to scenarios where network connectivity and computing capability extend to any physical objects and sensors,  allowing these devices to generate, exchange, and consume data with minimal human intervention. For the engineer, and in particular the communication and signal processing engineer for which the present article \textcolor{black}{is destined for}, IoT raises new technical challenges due to its special nature compared to conventional communication networks.  The requirements in terms of quality of service may be stringent e.g., in terms of  latency and reliability. IoT communications may involve devices having markedly different natures (humans and machines), devices with very different communication and computation capabilities. IoT communications may also involve exchanging massive data which could be multimodal and of large dimension. To address these challenges, the conventional communication engineering methodology, partly inherited from Shannon communication model, needs to be augmented. Goal-oriented (GO) communication paradigm, which is presented next, precisely corresponds to one of the novel communication approaches aiming to solve the aforementioned technical challenges posed by the deployment of the IoT. 

\textcolor{black}{While the design of conventional communication systems generally targets accuracy, fidelity, or reliability, \textcolor{black}{independently of the content}, the rationale behind GO communications is to adapt the encoded signal according to the specific, application-driven needs of the destination(s).} For instance, under the conventional communication paradigm, an IoT sensor may transmit an image of 1 Mbyte to a receiver whose goal/task is to decide about the absence/presence of a given feature in the image. This could be highly inefficient since the receiver only needs one bit of information and this bit could be directly sent by the transmitter. This toy example gives a rough idea about the potential benefits in terms of compression efficiency, spectral efficiency, or interference management in IoT networks. In a wireless sensor network,  it is of \textcolor{black}{high practical relevance to have} sensors that send the minimum amount of information required from the fusion center so as to take its decision properly and timely, avoiding unnecessary redundancies, which are potentially associated with computational and storage wastage, spectral efficiency losses, and network congestion. Similarly, for machine-to-machine (M2M) communications \cite{Mehmood-CM-2017}, the machine at the reception side might only need a given amount of information to execute a given task properly. These IoT applications clearly highlight the importance of tailoring the measured, encoded, and/or transmitted signal(s) to the final use(s) of it at the receiver(s).

The embryonic form of GO communications \textcolor{black}{was first discussed in \cite{Weaver}, where the semantic and effectiveness levels were introduced, and it was further developed in a more formal manner in \cite{Juba-jacm-2012} by proposing a mathematical model to mainly address the communication misunderstanding problem caused by imperfect knowledge of the protocol/language agreement.} However, this formal model is not suited to engineering problems, such as those encountered at the lowest layers of a communication system, the physical layer in particular. In contrast, the model we describe in this survey is based on a synthesis of recent research works {\cite{Walid,Stavrou-ISIT-2022,Barbarossa-CN-2021,Eldar-TSP1-2019,Zhang-AE-2021,Zou-JSAC-2022,tolga21SP}}, \textcolor{black}{which focus on} the lowest layers of a communication system, including key signal processing operations such as coding and estimation. In these works, both GO and semantic communications are considered. Semantic communication typically corresponds to scenarios in which the receiver is interested in the semantics, defined as either the meaning or the importance of the source message. The semantic aspect can be quantified through a measure of the message significance with respect to the final goal of the receiver, hence a natural connection between GO and semantic communications and the possibility to unify both approaches. In this respect, semantic communications in which efficiency/effectiveness of semantic transmission is explicitly defined and targeted can be qualified as goal-oriented communications in which the semantic nature of the information content or message is exploited.

In this survey, we show how key signal processing problems in IoT networks such as data compression, data clustering, data estimation, or machine learning are naturally related to the GO communication paradigm. For this purpose, the present survey is structured as follows. \textcolor{black}{In Sec. II, the approach of GO communications is described and its connections with semantic communications are discussed. In Sec. III, the focus is on one case study, namely GO data compression. It is clearly explained how some fundamental compression operations in IoT networks (signal preprocessing, quantization, and clustering) are revisited when tackled from the GO communication perspective. Open problems are discussed in Sec. IV. The survey is concluded by Sec. V.}

\section{\textcolor{black}{GO communication and its connections with semantic communication}}

To assess the potential of GO communications for IoT and understand how the GO paradigm is exploited in the IoT for data compression (Sec. III), it is essential to describe in detail in what GO and semantic communications consist in. The foundation of \textcolor{black}{GO communication} is the unification of data generation and transmission with the subsequent usage of the received information.  Specifically, extracting the relevant to the goal features of the transmit messages, characterizing the deviation of decision-making process induced by transmission noise and exploring the ultimate performance degradation brought by corrupted decisions, \textcolor{black}{GO communication} aims to build a new paradigm that could mitigate the adverse effect of transmission errors on the given goal.\textcolor{black}{Broadly speaking, semantic communication is the transmission of complex data structures (e.g., patterns, features, data lying on manifolds) or more generally of abstract concepts. As such, semantic communication is a broader concept than GO communication, since information semantics is not necessarily linked to an overarching system goal (letting aside a teleological explanation of physical processes).  Under this view, GO communications is a subset that provides a pragmatic view of semantic communications, in which the receiver is interested in the significance and the effectiveness (semantics) of the transported source message to achieve a certain task or goal.}


One of the purposes of the present section is to have a unified view of a set of recent works on GO and semantic communications \textcolor{black}{which are very relevant for IoT networks. For this purpose, we use a generic} model or structure for this type of communications. \textcolor{black}{The model is represented by Fig.~1. This figure is explained in details in the next subsections. To better grasp the practical implications behind these explanations, we will refer several times to the running example of GO data clustering for scheduling flexible power consumption in IoT networks where the sensors have: to monitor the non-flexible power consumption of some consumption devices or appliances (e.g., a network of smart meters); to cluster the measurements; to report the representatives of the clusters to a decision-making entity that controls flexible consumption (e.g., charging an electric vehicle). Several observations from Fig.~1 are in order.} 

First, a key observation is that the same terminology as \cite{Weaver} is used, in which communications can be categorized into \textbf{three levels}: the technical level, the semantic level, and the effectiveness (or efficiency) level. The conventional communication paradigm mainly focuses on the technical level to manage the communication reliability, whereas the meaning of the messages and their impact upon receipt is considered at the semantic and effectiveness levels, respectively. 

Second, the message generation and compression is presented at the source side. When agents are equipped with intelligence, an initial but important step consists in forming meaningful messages to serve its intended entity for reaching the goal. For this, the event occurred at the source needs to be observed, interpreted represented and transmitted by symbols. 

Third, the decision-making process concerning the effectiveness level is presented at the receiver side. It could be seen as an intermediate process to formulate the goal that depends on both states of source and the decision or actuation performed at the destination. Owing to the presence of decision-making process, the intent of the communication system is not only for bit and semantic reconstruction, but also for task accomplishment. In addition, it is worth  mentioning that the presence of these effectiveness level modules have an impact on the design of other levels modules since the information transmission techniques should be tailored to better serve the goal.

\begin{figure}[htp]
    \centering
    \includegraphics[width=9.5cm]{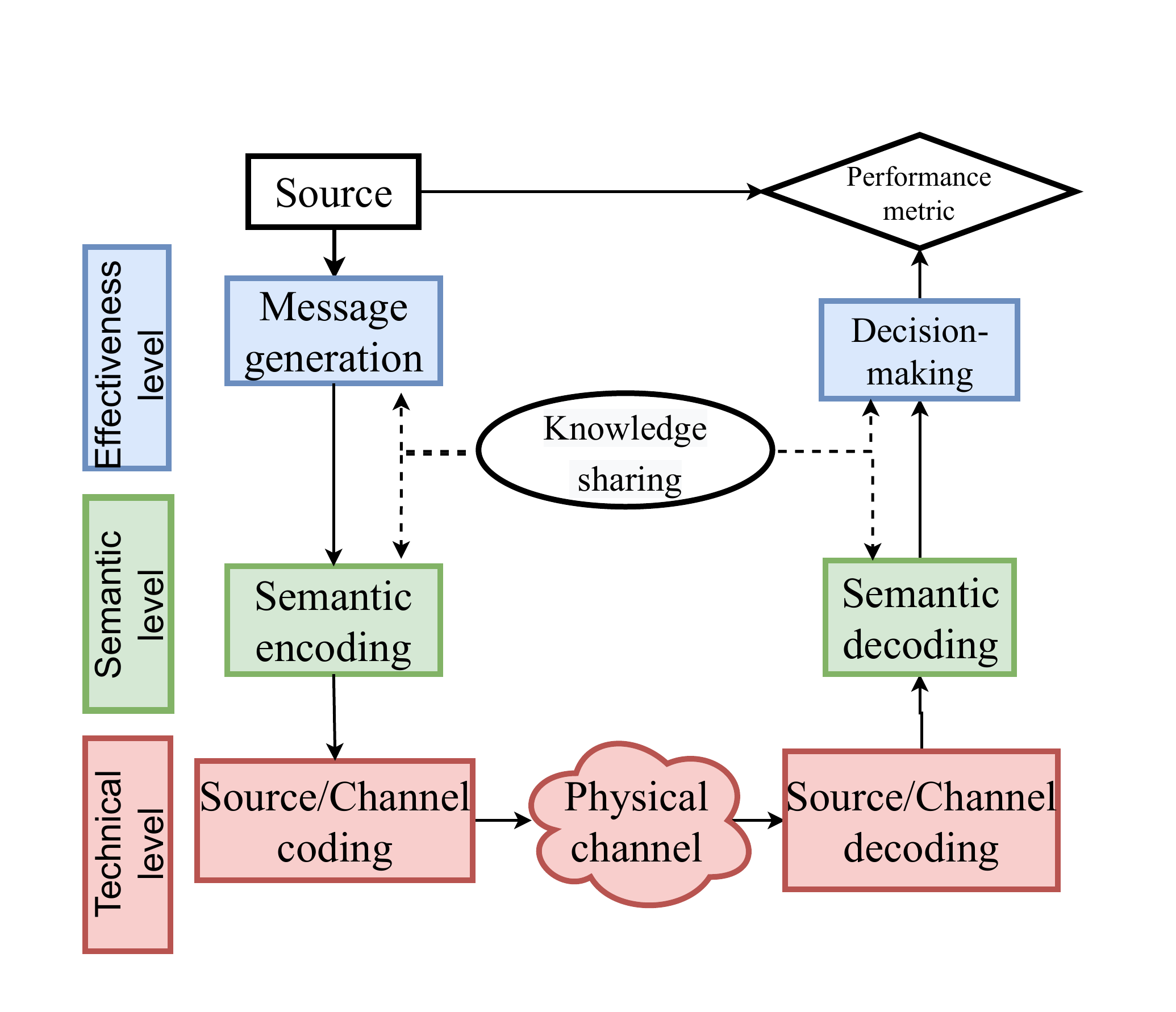}
    \caption{\textcolor{black}{A model that encompasses} goal-oriented and semantic communications}
    \label{fig-model}
\end{figure}

\subsection{A \textcolor{black}{generic} model for GO communications} 

In what follows, we comment on each module of Fig.~1.


\textbf{Message generation/conceptualization:} Unlike traditional communication systems, the information source for GO communication systems is not restricted to be an electronic device. Now agents with computing and \textcolor{black}{intelligence capabilities} (e.g., humans or AI-powered machines) may be typical sources, which leads to a more complex information generation step. Before the transmission process takes place, the prerequisite is to form the intended message to be sent. \textcolor{black}{When the entity/event of interest at the source cannot be made  identified or observed,  mapping methods or probabilistic models \cite{Stavrou-ISIT-2022} are often used to generate symbols.} Symbol representation can be seen as an interpretation of the entity/event and treated as the input of the following communication process.  \textcolor{black}{In the example of data clustering for the power consumption scheduling, the message generation corresponds to the knowledge of a database of non-flexible power consumption measurements that have collected by the sensors.}


\textbf{Semantic encoding/decoding:} In contrast to conventional communication systems, a semantic encoder is included in GO communication systems for efficient information representation. One common way in semantic encoding is to filter out irrelevant \textcolor{black}{parts} of the data to avoid unnecessary redundancy. By solely transmitting useful information that could impact the goal accomplishment, the censored data could achieve the same \textcolor{black}{(or approximately) performance as with the original data requiring lower rate.} Another typical technique in semantic encoding is to design a semantic-aware transformation scheme, which enables a GO sparse representation by mapping the data from a high dimensional space into a low dimensional space (e.g., information manifold) \textcolor{black}{often subject to a distortion.} Instead of completely removing some elements from the data, feature extraction approaches are applied here to obtain important attributes. Due to the success of deep learning methods in feature extraction functionality of several computer science tasks (e.g., nature language processing, computer vision), neural network based semantic encoders \cite{Xie-TSP-2021} are envisioned to become widely implemented. \textcolor{black}{In the above-mentioned power consumption scheduling problems, the semantic encoder is to extract features (e.g., the peak power occurrence time-slot) of the consumption profiles which are relevant for the peak power minimization. The semantic encoder determines how the consumption profile will be divided.}

\textcolor{black}{Furthermore, we notice} that a semantic encoder can be jointly designed with the source encoder. Classical source encoders typically exploit some knowledge of the probability distribution of the signal/symbols to be encoded. For instance, one typically allocates more bits to events/symbols having higher probabilities of occurrence. This approach is used for lossy source coding (see e.g., the case of scalar quantization in the high resolution regime \cite{Gray_TIT_1998}). However, when the signal reconstruction \textcolor{black}{in its entirety is not the ultimate purpose}, a much more efficient semantic-aware source encoder can be used. New quantities stemming from an appropriate combination of these key factors could be a solution, where a weighted distribution encompassing the original distribution and the derivatives of the final utility function has shown its optimality in asymptotic cases \cite{Zou-JSAC-2022}. 


\textbf{Knowledge bases:} In GO communication systems, the amount of information conveyed by a message is determined by both the message itself and also the level of knowledge available at the transmitter and the receiver. The encoding process can exploit the knowledge base system to maximize the information conveyed by the message (where the semantics/meaning could be recovered from the receiver knowledge base) and minimize the amount of data to be transmitted. Information
with higher inference difficulty should be encoded more precisely while information that is simple to interpret can be encoded coarsely. 

\textbf{Decision-making:} A salient feature of GO communications stems from the combination of decision-making and transmission processes. \textcolor{black}{For the example of GO clustering performed by the sensors, the decision is taken by a central entity and consists in choosing flexible power consumption profiles based on the knowledge of the representatives reported by the sensors. In the example of a wireless communication network, the monitored information may be the channel state information and the decision may be the radio resource allocation policy (e.g., a beamforming vector or a precoding matrix).} In general, most decision making processes could be divided into two groups. First, when the decision making process can be modeled as an optimization problem (OP) where the form of the objective function and constraints could be explicitly expressed.  Second, when the mapping corresponding to the decision making process cannot be directly interpreted, but one still has access to the output for a given input. This scenario occurs often in computer science problems, such as object recognition. Deep learning architectures are often implemented to explore the mapping under the supervision of given decision results. 


\subsection{Performance metrics}

Under the GO paradigm, what matters is the degree of goal fulfillment. This means that the retained performance metric should be semantic-aware and goal-oriented. To understand the difference between the conventional metrics and metrics used for GO communications, we discuss below three types of performance metrics used to evaluate the performance of a GO communication. 

\textbf{Conventional distance metrics:} Conventional metrics mainly aim to evaluate the similarity between the transmitted symbols and received symbols. Distance metrics such as the Euclidean distance or Hamming distance are used to measure the difference between the original and the reconstructed information. However, these distances measure the transmission accuracy regardless of the subsequent processes and the ultimate goal of the communication and could lead to an inefficient use of communication resources.

\textbf{Semantic metrics:} \textcolor{black}{In the semantic level, the used metrics have to ensure the transmission accuracy and logical truth of the semantic information.} Even though errors could be observed in the physical channel transmission, the information retrieved by the semantic decoder could be lossless. As shown in Fig.~1, the semantic metric can measure the semantic similarity between input symbols of the semantic encoder and output symbols of the semantic decoder. For text transmission, the semantic metric could be word similarity or sentence similarity. For instance, cross entropy between the word probability distribution in real sentences and the word probability distribution in reconstructed sentences is used as the loss function in DeepSC \cite{Xie-TSP-2021}.  In other type of applications, the peak signal-to-noise ratio (PSNR) and signal-to-distortion ratio (SDR) can be seen as semantic metrics in the image transmission and speech transmission, respectively.

\textbf{GO metrics:} \textcolor{black}{In the GO paradigm, the task to be executed is usually modeled by an optimization problem (OP).} \textcolor{black}{The function to be optimized is assumed to have two types of variables: the decision variables associated with the task to be performed; the function parameters that precisely correspond to the information which is encoded. A natural and fundamental question is then: How to compress the function parameter information to achieve a given performance loss w.r.t. the ideal case of perfect information? Answering this key question falls into the problem of GO compression.} \textcolor{black}{In the case of data clustering for power consumption scheduling, the goal/task may be typically to minimize the peak power of the total consumption profile. The semantic/feature extraction associated with clustering should not be chosen from exogenous indices or criteria but tailored to this objective.} The goal function may typically have decision variables and source-dependent variables. Another simple example of source-dependent variables in wireless communications is given by the CSI. For instance, the final metric may correspond to spectrum efficiency maximization, which is dependent of the channel state. Rather than solely consider the transmission accuracy of transmitted messages or consider the quality of key features (or semantics) in reception, GO metrics could be seen as combined metrics that jointly evaluate the transmission errors in physical channels, the extracted features transmission accuracy, and the ambiguity in communication processes to the goal. Ideally, when the system goal could be explicitly represented by a utility function, the GO metric could be the utility function itself or one of its variants. If the goal cannot be formulated by a given utility function, metrics could be quantities representing the degree of goal accomplishment. Even though the mathematical form of the utility function is unknown, providing the goal accomplishment results could stimulate the message shaping to better serve the goal. Supervised learning could enhance the performance of interest in an offline manner by using goal-dependent labels, and reinforcement learning could further improve the performance in an online manner with goal-dependent rewards. 

\section {Goal-oriented data compression in the IoT}
%


To deal with the massive presence of sensing devices in IoT networks and also to manage the available resources such as the radio spectrum, computation, and storage capabilities efficiently, data compression \textcolor{black}{constitutes} a key signal processing operation for the IoT. As explained in the previous sections, considering distortion-like or distance-based performance metrics \textcolor{black}{at the technical level} generally induces an inefficient \textcolor{black}{use} of communication resources because they are chosen independently of the \textcolor{black}{final use of the information.} By contrast, if only the semantic of the compressed data has to be recovered at the receiver or if the receiver has only a precise task to execute, the usage of resources may be dramatically improved. Accounting for the presence of a goal through a given utility function requires to revisit the classical compression operations. These operations include for instance, signal preprocessing , signal quantization, \textcolor{black}{lossless source coding (e.g., entropy coding),} segmentation, block matching, object tracking, flow detection, feature extraction, or data clustering. \textcolor{black}{In what follows,} the focus is made on four types of operations, namely signal preprocessing, quantization, clustering, and \textcolor{black}{causal reasoning based compression}, and for each of them, the impact and benefits of the \textcolor{black}{GO communication} paradigm are discussed. 


\subsection{Goal-oriented signal preprocessing}

\textcolor{black}{Signal preprocessing is common in data compression systems. It typically serves as a preparation for the compression operations that take place just after the preprocessing block. Preprocessing may e.g., consist in filtering to de-noise the input signal, in applying a transform to change the working signal domain or in reducing the complexity of the subsequent operations.} An interesting GO compression approach is proposed in \cite{Eldar-TSP1-2019} which combines linear processing and uniform quantization. The input signal first undergoes a linear pre-processing module (a matrix) and is transformed into a lower \textcolor{black}{dimensionality signal}. What is interesting is that the choice of the linear transformation is tailored to the final utility (or task efficiency) function. For instance, for a given distortion level, the best linear transformation in terms of rank reduction is known to be the Karhuenen-Loeve transform (KLT) but this result does nor hold anymore in the presence of a general utility function. Also, it is known that a linear transformation is optimal for Gaussian inputs, but again this result is no longer valid in the presence of a general goal function, which motivates the use of non-linear transformation. For this purpose, the authors of \cite{Eldar-Entropy-2021} propose to learn such a non-linear transformation by using a deep neural network (DNN). \textcolor{black}{In IoT networks such as those involving sensors to monitor the energy consumption of the consuming devices, a typical task for the whole system would typically be to minimize the total peak power. For this purpose, in \cite{Sun-WINCOM-2019}, the linear and non-linear transformation (applied to the non-flexible power vector) that minimize the Lp-norm of the sum of the non-flexible power consumption vector and the flexible power consumption vector is determined.} Compared to the KLT in the presence of real smart grid data from the Ausgrid database, very significant gains are obtained in terms of optimality loss induced by the rank reduction \textcolor{black}{when using these transformations}.  
\subsection{Goal-oriented quantization} 

\begin{figure}[htbp]
\centering{}\includegraphics[scale=0.42]{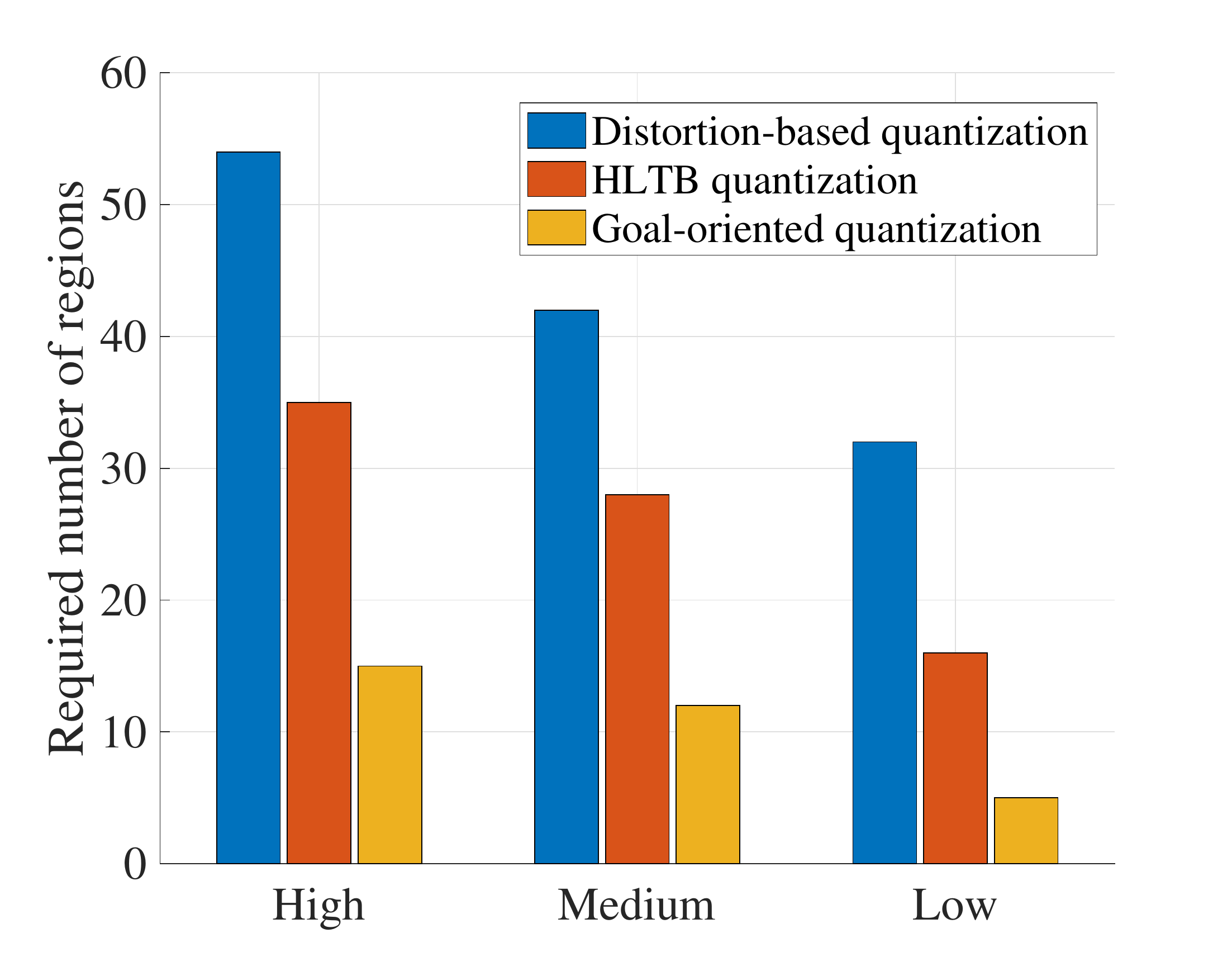}
\caption{\textcolor{black}{Required number of quantization regions to achieve a given relative optimality loss target (low for $20\%$, medium for $5\%$ and high for $1\%$) for GO, HLTB and distortion-based approaches. The large gains by implementing goal-oriented quantization can be clearly observed in all precision levels. }\label{fig:GOQ_vs_DB}}
\end{figure}

Quantization is often a fundamental element of a data compression system. Indeed, at least under the classical communication paradigm, it necessarily induces an information loss and constitutes a natural degree of freedom to tune the compression rate of the source coder. Just as for the signal preprocessing stage, known results have to be revisited to avoid potentially large efficiency losses. \textcolor{black}{To be specific, it is known that a good quantizer (in the sense of the distortion) needs to be adapted to the quantizer input signal. In some cases, like high-resolution scalar quantization, the formal link between the quantization function and the input distribution can be fully determined. However, for a general goal function, the quantization rule has to take into account the smoothness and regularity properties of the goal function and the decision function associated with the task. This has been formally proved for vector quantizers in the high resolution regime \cite{Zou-JSAC-2022}.} In particular, it is shown that it is no longer optimal to allocate more bits to the most likely signal realizations or input symbols but rather according to a law which results from the effect of the input distribution and the decision function properties. Additionally, the impact of the goal on the compression rate is studied, which allows \textcolor{black}{one} to constitute classes of functions for which GO compression yields very large gains or moderate gains. \textcolor{black}{In contrast with conventional distortion-based quantization and hardware limited task-based (HLTB) quantization presented in \cite{Eldar-TSP1-2019}, \textcolor{black}{Fig. \ref{fig:GOQ_vs_DB}} shows the potential of GO quantization in saving communication resources for a given quadratic utility function. }

\subsection{Goal-oriented data clustering}
Here, we consider \textcolor{black}{scenarios in which a database of signal samples is available; for instance, in networks with sensors monitoring power consumption (e.g., smart meters), a sequence of power levels recorded over one day would typically constitute a sample. From this database, groups of data or clusters has to be formed and possibly represented by a representative sample. The idea of GO clustering is to adapt these groups and possible representatives to the final task that has only access to the clustered data (say a representative) and not to the original data. In the example of power consumption, the sensor would only report to the decision-making entity the closest representative consumption vector and not the actual measurements. Here also,} classical results known for data clustering are no longer valid in the presence of a general goal function.  Using ``Voronoi-type" data clusters \textcolor{black}{(as the k-means clustering technique would provide)} may be very inefficient as well. This has been shown in \cite{Zhang-AE-2021} where the authors consider the $L_{\infty}$-norm for the goal function and derive the optimal clusters given some representative choices, and vice versa. To illustrate the gain brought by GO clustering \textcolor{black}{for this application}, GO clustering  \cite{Zhang-AE-2021} has been compared to known clustering schemes such as the conventional k-means algorithm and hierarchical clustering. \textcolor{black}{Fig.~4 illustrates the dramatic reduction in terms of required number of clusters to achieve a given optimality loss (w.r.t. the case where the data would be perfectly known to optimize the goal function) for the goal function.}




\begin{figure}[htbp]
\centering{}\includegraphics[scale=0.40]{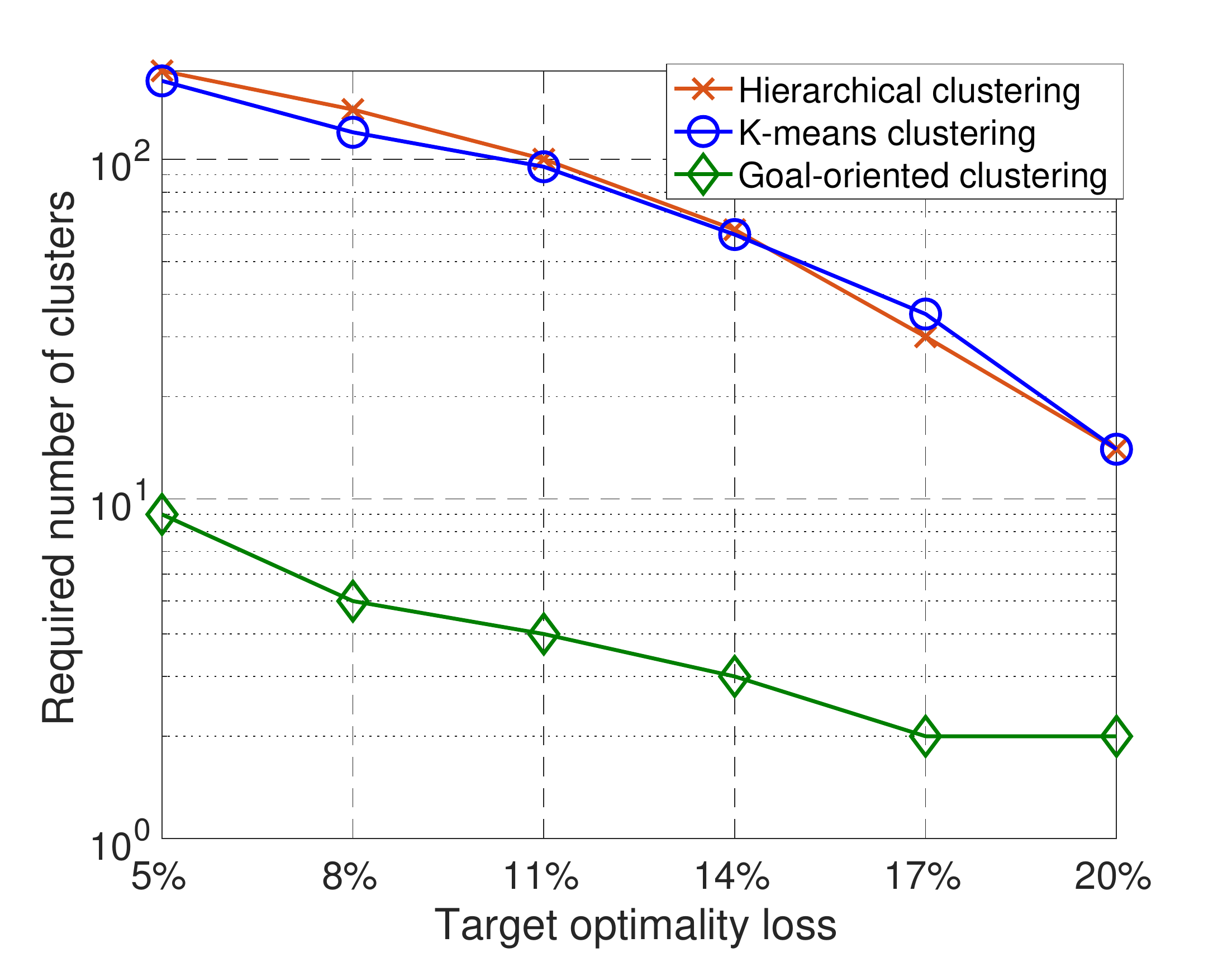}
\caption{\textcolor{black}{Required number of clusters to reach a target optimality loss for peak power minimization problems. A significant gap between GO clustering and conventional clustering methods can be observed. While the conventional clustering schemes require fifteen clusters to reduce the optimality loss below 20\%, GO clustering method solely needs two clusters to achieve the same goal. }} \label{fig:GOQ-OP-MU}
\end{figure}

\begin{figure}[htbp]
\centering{}\includegraphics[width=0.48\textwidth]{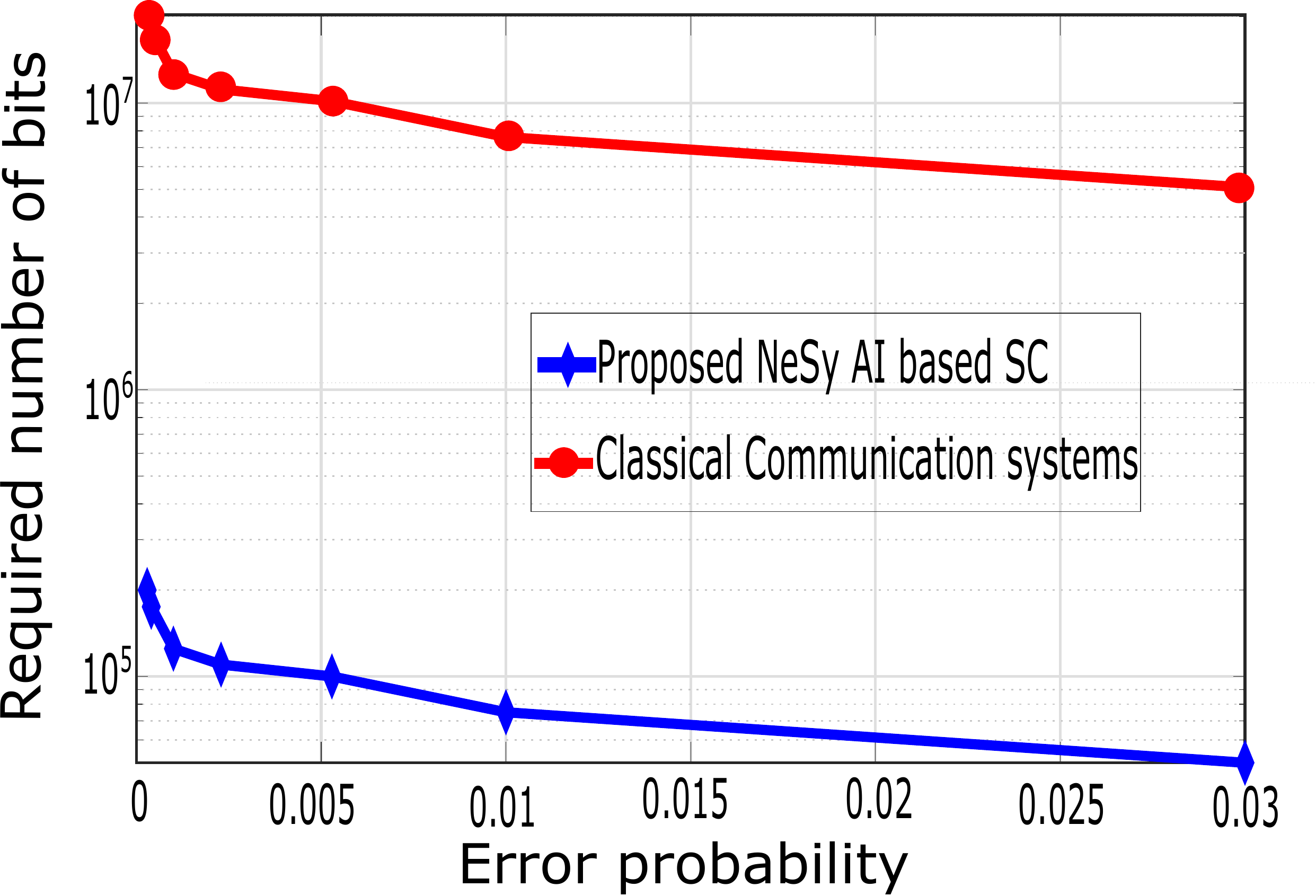}
\caption{\textcolor{black}{By combining reasoning tools with machine learning, the  devices can sense and communicate efficiently, by not sending large unnecessary data \cite{Walid}. }} \label{fig:walid-NeSy}
\end{figure}

\textcolor{black}{\subsection{Intent-based semantic communications}}

\textcolor{black}{Intent-based networks are autonomous systems that define the behavior they expect from their network, e.g., ``improving network quality,'' for the system to then automatically translate it into real-time network action. In such a context, integrating the semantic and effectiveness aspects is quite apropos. This requires the transmit and receive nodes to move away from being just blind devices (that transfer data back and forth) towards brain-like devices capable to understand and reason over the data and how it gets generated. Reasoning here implies enabling the nodes to make logical conclusions and generalizations out of the data. Motivated by this approach, the framework of neuro-symbolic (NeSy) artificial intelligence was proposed in \cite{Walid} to learn the causal structure behind the observed data and, then, use it to enhance semantic communications. This approach leads to significant gains in terms of transmission efficiency as shown in Fig.~\ref{fig:walid-NeSy}.}

\section{\textcolor{black}{Challenges and Open Problems}}
In this section, we discuss several open issues and technical challenges associated with \textcolor{black}{GO communication}.

\textbf{\textcolor{black}{GO communication} with time-evolving goals:} In systems with smart devices, it is common that one task is followed by another task(s). Once the ongoing task is completed, a new task need to be executed seamlessly. However, redesigning or retraining from scratch is not only time-consuming, but \textcolor{black}{also wasteful in terms of resources} (e.g., energy). In this respect, a unified GO communication framework that \textcolor{black}{accounts for multiple - often causally related - goals and maximizes the expected goal accomplishment could be a first solution when these intermediate goals are known in advance.} Moreover, an adaptive approach or transfer learning based approach based on previous model could be embedded in the framework when future goals are unknown.


\textbf{\textcolor{black}{GO} coding and control:} \textcolor{black}{ To characterize the goal-oriented compression, source coding  or joint source channel coding models could be implemented to study its limiting performance.} When the goal depends not only on the state and the decision taken in the current timeslot, but also on previous timeslots, dynamic system models \textcolor{black}{could be a better formulation.} In this case, using a simple objective function or learning a mapping method through neural networks might turn out to be insufficient. One possible solution is to resort to differential equations and explore the connections and the evolution between transmitted messages and the goal. \textcolor{black}{Another aspect that has to be revisited is the sampling process. There already exist many technical contributions dedicated to event-triggered sampling, which the objective being to attain dynamical system stability.} It seems very relevant to tailor the sampling problem to a general utility function just as it has been done for the quantization problem.

\section{Conclusion}
\textcolor{black}{A highly promising technical solution to face challenges stemming from the deployment of IoT devices, such as high device density, stringent quality of service requirements, and massive, multimodal,} and heterogeneous data, is to resort to goal-oriented and semantic communications. This paper provides a comprehensive overview of goal-oriented and semantic communications. When communication is not an end \textcolor{black}{but a means to achieve an ultimate goal, data semantics should be taken into account in the design of future communication systems, in particular when the goal deviates from entire signal reconstruction.} Exploring a data compression case study, which is a foundational problem in IoT systems, we have discussed the potential benefits from deploying GO compressors and GO signal processors in general. Our conclusions hold for the signal processing part, quantization, and data clustering, but the mentioned insights may have a much wider applicability that concerns many other signal processing operations in IoT communication networks.






\end{document}